\documentclass[final]{aipproc}
\layoutstyle{6x9}

\def\bra{\langle}
\def\ket{\rangle}
\def\qu{\langle\bar u u\rangle}
\def\qd{\langle\bar d d\rangle}
\def\qs{\langle\bar s s\rangle}
\def\qq{\langle\bar q q\rangle}
\begin{document}

\title
{ Unitary Structure of the
QCD Sum Rules and 
$K Y N$ and $K Y \Xi$ Couplings
}



\author {T.~Aliev}{
address={Physics Dept., Middle East Technical Univ., 06531, Ankara, 
Turkey},
email={taliev@metu.edu.tr}}

\author {A.\"{O}zpineci}{
address={INFN, Sezione di Bari, Bari, Italy
\thanks{Present address: Middle East Technical University, Ankara,Turkey}},
email={ozpineci@ba.infn.it}}

\author{S.B.~Yakovlev}{
address={D.V.~Skobeltsyn Institute of Nuclear Physics,
Moscow State University, Moscow, Russia},
email={zamir@depni.sinp.msu.ru}}

\author{V.S.~Zamiralov}{
address={D.V.~Skobeltsyn Institute of Nuclear Physics,
Moscow State University, Moscow, Russia}}

\begin{abstract}
New relations between QCD Borel sum rules for 
strong coupling constants of K-mesons to baryons are derived.
It is shown that starting from the sum rule for the
coupling constants $g_{\pi\Sigma\Sigma}$ 
and  $g_{\pi\Sigma\Lambda}$
it is straightforward to obtain
corresponding sum rules for the $g_{K Y N}$,
$g_{ K Y \Xi}$ couplings, $Y=\Sigma,\Lambda$.
\end{abstract}

\maketitle
\section{Introduction}

Meson-baryon couplings were studied for years
thoroughly either for pion-baryon couplings or
kaon-baryon baryon ones as these couplings are important 
parameters of strong interaction physics.

Since the advent of the $SU(3)$ symmetry 
all the meson-baryon coupling constants were usually expressed in terms
of $F$ and $D$ constants which gave possibility to construct
a reliable  phenomenological aproach. 

As soon as in \cite{Shif} QCD sum rules (SR's) were proposed, they 
were used not only for baryon masses and magnetic moments
starting from the works \cite{Ioffe}
but also for baryon-meson coupling constants. Naturally, 
a pion-nucleon coupling attracted the most attention
(see, for example, \cite{Sciomi}, \cite{Krippa}, \cite{Kim0}).
Coupling constants of $\pi^{0}$- and $\eta$- mesons
to baryons were studied recently in various QCD SR approaches
\cite{Kim}, 
\cite{AOS}. Also QCD sum rule for the  $\eta$ coupling to the
$\Lambda$ hyperon was written \cite{Oyzslpi} which 
was usually absent in these approaches. 

As for K-mesons they were also
studied in the framework of the QCD sum rules ( see, e.g.,
\cite{Choe1}, \cite{Choe2}, \cite{Marina}). Usually these sum rules
are not related straightforwardly to those treating
$\pi$ and $\eta$ couplings to baryons.

We would like to propose here QCD sum rules for octet
meson-baryon couplings through some universal
${\cal{F}}$ and ${\cal{D}}$ functions written in a unified manner.
In order to be clear we choose for methodical reasons as a basis for our
reasoning $SU(3)$ breaking QCD Borel sum rules proposed in \cite{Kim}.

\section{Relation between $\pi^{0} \Sigma\Sigma$ 
and  $\pi\Sigma\Lambda$ constants in SU(3)}

We begin as in \cite{OYZ} with a simple example. 
In the unitary model all the pion-baryon coupling constants 
can be expressed in terms of $F$ and $D$ coupling constants.

But coupling of the $\Sigma$-like baryons $B(qq,q^{\prime}), 
q,q^{\prime}=u,d,s$ to $\pi^{0} $ meson
related in the quark model to the current
$j^{\pi^{0}}=\frac{1}{\sqrt{2}}
[ \bar u \gamma_{5}u-\bar d \gamma_{5}d]$
can be put in the form
$$
g(\pi^{0} B B)==g_{\pi qq}2F+g_{\pi q^{\prime}q^{\prime}}(F-D),
$$
or, particle per particle:
$$
g(\pi^{0} p p)=g_{\pi uu}2F+g_{\pi dd}(F-D)=\sqrt{\frac{1}{2}}(F+D);
$$
$$
g(\pi^{0} \Sigma^{+} \Sigma^{+})=
g_{\pi uu}2F+g_{\pi ss}(F-D)=\sqrt{2}F,
$$
and so on, 
where $g_{\pi uu}=+\sqrt{\frac{1}{2}}$ ,
$g_{\pi dd}=-\sqrt{\frac{1}{2}}$ and    
$g_{\pi ss}=0$ are just read off the quark current.

The only coupling which cannot be written
immediately in this way is $\pi^{0}\Sigma^{0}\Lambda$.
To overcome this difficulty
let us write for $\pi^{0}\Sigma^{0}\Sigma^{0}$
coupling ( which is equal to zero !):
\begin{equation}
g(\pi^{0} \Sigma^{0} \Sigma^{0})=
g_{\pi^{0} uu}F+g_{\pi^{0} dd}F+g_{\pi^{0} ss}(F-D)=0
\label{su3Si0}
\end{equation}
and change ($d\leftrightarrow s$) and  ($u\leftrightarrow s$)
to form two auxiliary quantities 
\begin{equation}
g(\pi^{0} \tilde\Sigma^{0,ds} \tilde\Sigma^{0,ds})=
g_{\pi^{0} uu}F+g_{\pi^{0} ss}F+g_{\pi^{0} dd}(F-D)=\sqrt{\frac{1}{2}}D,
\label{su3Sids0}
\end{equation}
\begin{equation}
g(\pi^{0} \tilde\Sigma^{0,us} \tilde\Sigma^{0,us})=
g_{\pi^{0} dd}F+g_{\pi^{0} ss}F+g_{\pi^{0} uu}(F-D)=-\sqrt{\frac{1}{2}}D.
\label{su3Sius0}
\end{equation}
The following relation holds:
\begin{equation}
g(\pi^{0} \tilde\Sigma^{0,ds} \tilde\Sigma^{0,ds})-
g(\pi^{0} \tilde\Sigma^{0,us} \tilde\Sigma^{0,us})
=\sqrt{3}g(\pi^{0} \Sigma^{0} \Lambda).
\label{silapi}
\end{equation}
The origin of this relation lies in the structure of
$\Sigma^{0}(ud,s)$ and $\Lambda$ wave functions in the NRQM.
With the exchanges $d\leftrightarrow s$ and $u\leftrightarrow s$ 
one arrives at the corresponding $U$-spin and $V$-spin quantities, so
\begin{eqnarray}
\left(\begin{array}{cc}
-\tilde\Sigma^{0}_{ds}\\
\tilde\Lambda_{ds}
\end{array}
\right)=
\left(\begin{array}{cc}
1/2&\sqrt{3}/2 \\
-\sqrt{3}/2&1/2
\end{array}
\right)
\left(\begin{array}{cc}
\Sigma^{0}\\
\Lambda
\end{array}
\right),
\left(\begin{array}{cc}
-\tilde\Sigma^{0}_{us}\\
\tilde\Lambda_{us}
\end{array}
\right)=
\left(\begin{array}{cc}
1/2 & -\sqrt{3}/2 \\
\sqrt{3}/2& 1/2
\end{array}
\right)
\left(\begin{array}{cc}
\Sigma^{0}\\
\Lambda
\end{array}
\right).
\label{Lads}
\end{eqnarray}
It is easy now to show that the relation  Eq.(\ref{silapi}) follows
and it shows us the way to proceed with the QCD sum rules.

\section{$K Y N$,
$KY\Xi$ and  $\pi\Sigma\Lambda$ couplings in the SU(3)}

Now we consider  kaon and charged pion couplings
to baryons. They are given by $SU(3)$ symmetry formulae but
we rewrite it in a way suitable for derivation of
the corresponding Borel sum rules.
Let us write coupling of pion to $\Sigma^{+}$
and $\Lambda_{ds}$ given by the Eq.(\ref{Lads}):
\begin{eqnarray}
2[g(\pi^{-} \Sigma^{+}\bar \Lambda_{ds})]=
-\sqrt{3} g(\pi^{-} \Sigma^{+}{\bar \Sigma}^{0})
+g(\pi^{-} \Sigma^{+}\bar \Lambda)=
\end{eqnarray}
$$
-\sqrt{3}(-\sqrt{2} F) + \sqrt{\frac{2}{3}}D= 
\sqrt{\frac{2}{3}}(3F+D).
$$
Now we perform $d\leftrightarrow s$ exchange.
Our auxiliary baryon $\Lambda_{ds}$ returns to
real $\Lambda$ while $\pi^{-}(\bar d u)$ changes to
$K^{-}(\bar s u)$ and $\Sigma^{+}(uu,s)$ changes to $-p(uu,d)$,
so that
\begin{eqnarray}
2[g(\pi^{-} \Sigma^{+} \Lambda_{ds})]_{ds}=
-2[g(K^{-} p \Lambda )]=\sqrt{\frac{2}{3}}(3F+D).
\end{eqnarray}
This is the unitary symmetry result.
In the same way we write the formal coupling of  pion to $\Sigma^{+}$
and $\Lambda_{us}$ given by the Eq.(\ref{Lads})
and then perform $u\leftrightarrow s$ exchange to obtain
\begin{eqnarray}
2[g(\pi^{-} \Sigma^{+}\bar \Lambda_{us})]_{us}=
2[g(K^{0} \Xi^{0} \bar \Lambda )]=-\sqrt{\frac{2}{3}}(3F-D).
\end{eqnarray}
This is again the unitary symmetry result.
Similarly one can show that 
\begin{eqnarray}
-2[g(\pi^{-} \Sigma^{+}\bar \Sigma^{0}_{ds})]_{ds}=
2[g(K^{-} p \bar \Sigma^{0} )]=\sqrt{2}(-F+D),
\end{eqnarray}
$$
-2[g(\pi^{-} \Sigma^{+}\bar \Sigma^{0}_{us})]_{us}=
2[g(K^{0} \Xi^{0} \bar \Sigma^{0} )]=-\sqrt{2}(F+D).
$$
Derivation of these coupling constants indicates us the way to
proceed in the formalism of QCD sum rules.

\section{QCD sum rules}

We use as the example QCD sum rules based on the formalism
developed in \cite{Kim} where unitary symmetry is broken but
formules are rather transparent. The sum rule for the 
${\cal M} \Sigma^{0}\Sigma^{0}$ coupling reads:
\begin{eqnarray}
\frac{1}{\sqrt{2}}m_{{\cal M}}^{2}\lambda^{2}_{\Sigma}
g({\cal M} \Sigma^{0}\Sigma^{0})e^{-(m_{\Sigma}^{2}/M^{2})}
[1+A_{\Sigma}M^{2}]=
\nonumber\\
g_{{\cal M} ss}m_{{\cal M}}^{2}M^{4}E_{0}(x)[\frac{\qs}
{12 \pi^{2}f_{{\cal M}}}+\frac{3f_{3{\cal M}}}{4\sqrt{2}\pi^{2}}]
\nonumber\\
-g_{{\cal M} ss}\frac{1}{f_{{\cal M}}}M^{2}
(m_{d}\qu+m_{u}\qd)\qs
\nonumber\\
-g_{{\cal M} ss}\frac{m_{{\cal M}}^{2}}{72f_{{\cal M}}}\qs
\bra \frac{\alpha_{s}}{\pi}
{\cal{G}}^{2}\ket
\nonumber\\
+\frac{1}{6f_{{\cal M}}}m_{0}^{2}[\qs(m_{d}g_{{\cal M} uu}\qu
+m_{u}g_{{\cal M} dd}\qd)
\nonumber\\
+m_{s}(g_{{\cal M} uu}+g_{{\cal M} dd})\qu\qd].
\label{si0}
\end{eqnarray}
where $m_{q},\quad q=u,d,s $ are current quark masses, 
$f_{{\cal M}}$ is a ${\cal M}$-meson decay constant, 
${\cal M}=\pi^{0},\eta$, quark condensates are
$\qu=\qd=-(0.23)^3$ GeV$^{3}$, $\qs/\qd=0.8$, 
while $m_{0}^{2}=0.8$ GeV$^{2}$,
$
\bra\bar g_{c} q\sigma\cdot G q \ket\equiv \ket m_{0}^{2}\qq.
$
The factor $E_{0}(x)=(1-e^{-x})$ is used to
subtract  the continuum contribution, $x=W^2/M^2$ \cite{Ioffe}
(we take $W^2=2.0$ GeV$^{2}$). The overlap amplitude is taken 
as $\lambda^{2}_{B}=C\cdot M_B^6$ GeV$^{6}$ \cite{Kim},
with $C=5.48\times 10^{-4}$. We neglect in calculations
$f_{3{\cal M}}$. Parameter $A_{B}$ accounts for
high-resonance contributions.

We define ${\cal{D}}^{(0)}({\cal{M}};M^{2};u,d;s)$ and
${\cal{F}}^{(0)}({\cal{M}};M^{2};u,d;s)$
(this shorthanded notation means that they depend
on $M^{2}$, all quark masses and all condensates:
${\cal{D}}^{(0)}({\cal{M}};M^{2};u,d;s)$ $\equiv$ 
${\cal{D}}^{(0)}
({\cal{M}};M^{2};m_{u},\qu,...; m_{d},\qd,...; m_{s},\qs,...)$,
similar for ${\cal{F}}$
):
\begin{eqnarray}
{\cal{F}}^{(0)}({\cal{M}};M^{2};u,d;s)=
\frac{1}{6f_{{\cal{M}}}}m_{0}^{2}
[\qs(m_{d}\qu+m_{s}\qu\qd],
\nonumber\\
{\cal{D}}^{(0)}({\cal{M}};M^{2};u,d;s)-
{\cal{F}}^{(0)}({\cal{M}};M^{2};u,d;s)=-
[m_{{\cal{M}} }^{2}M^{4}E_{0}(x)
[\frac{\qs}
{12 \pi^{2}f_{ {\cal{M}} } }+\frac{3f_{3{\cal{M}}}}{4\sqrt{2}\pi^{2}}]
\nonumber\\
-\frac{1}{f_{{\cal{M}}}}M^{2}
(m_{d}\qu+m_{u}\qd)\qs
-\frac{m_{{\cal{M}}}^{2}}{72f_{{\cal{M}}}}\qs
\bra \frac{\alpha_{s}}{\pi}
{\cal{G}}^{2} \ket],
\label{D}
\end{eqnarray}
The righthand side (RHS) of the Eq.(\ref{si0}) can be written in a form
\begin{eqnarray}
RHS({\cal{M}} \Sigma^{0}\Sigma^{0})=
g_{{\cal{M}}uu}{\cal{F}}^{0}({\cal{M}};M^{2};u,d;s)+
g_{{\cal{M}}dd}{\cal{F}}^{0}({\cal{M}};M^{2};d,u;s)+
\nonumber\\
\frac{1}{2}g_{{\cal{M}} ss}({\cal{F}}^{0}({\cal{M}};M^{2};s,d;u)
+{\cal{F}}^{0}({\cal{M}};M^{2};s,u;d))-
\nonumber\\
\frac{1}{2}g_{{\cal{M}} ss}({\cal{D}}^{0}({\cal{M}};M^{2};u,d;s)+
{\cal{D}}^{0}({\cal{M}};M^{2};d,u;s) ).
\label{Sigma0fd}
\end{eqnarray}
With isotopic invariance we construct Borel sum rule for 
the $\pi^{-} \Sigma^{+}{\bar \Sigma}^{0}$:
\begin{eqnarray}
-m_{\pi}^{2}\lambda^{2}_{\Sigma}
g(\pi^{-} \Sigma^{+}\Sigma^{0})e^{-(m_{\Sigma}^{2}/M^{2})}
[1+A_{\Sigma}M^{2}]=
\nonumber\\
\frac{m_{0}^{2}}{6f_{\pi}}[ (m_{u}\qs+m_{s}\qu) (\qu+\qd)]\equiv
\sqrt{2}{\cal{F}}^{(-)}(\pi^{-};M^{2};u,d;s)
\label{pimisi}
\end{eqnarray}
and a similar sum rule for $\pi^{+} \Sigma^{-}\Sigma^{0}$
coupling (upon $ u \leftrightarrow d $).

Using analogue of the Eq.(\ref{silapi})
$$
\sqrt{3}RHS(\pi^{0}\Sigma^{0}\Lambda)=
RHS(\pi^{0}\Sigma^{0}_{ds}\Sigma^{0}_{ds})-
RHS(\pi^{0}\Sigma^{0}_{us}\Sigma^{0}_{us})
$$
we construct QCD Borel sum rule for
$\pi^{0}\Sigma\Lambda$ coupling \cite{Oyzslpi}
\begin{eqnarray}
\sqrt{3}m_{\pi}^{2}\lambda_{\Lambda}\lambda_{\Sigma}
 g(\pi^{0}\Sigma^{0}\Lambda) 
\frac{M^{2}}{M_{\Sigma}^{2}-M_{\Lambda}^{2}}
(e^{-M^{2}_{\Lambda}/M^{2}}-e^{-M_{\Sigma}^{2}/M^{2}})
[1+A_{\Sigma\Lambda}M^{2}]=
\nonumber\\
-m_{\pi }^{2}M^{4}E_{0}(x)[\frac{ \qd + \qu}
{12 \pi^{2}f_{ \pi  } }
+\frac{3f_{3\pi }}{4\sqrt{2}\pi^{2}}]
+\frac{m_{\pi }^{2}}{72f_{\pi }}[\qd + \qu ]
\bra \frac{\alpha_{s}}{\pi}
{\cal{G}}^{2} \ket
\nonumber\\
+\frac{1}{6f_{\pi }}(6M^{2}+m_{0}^{2})[
(m_s \qu+m_u \qs)\qd  +(m_d \qs+m_s \qd)\qu  ]
\label{lafin}
\end{eqnarray}
The RHS of it with Eq.(\ref{Sigma0fd}) can be put in the form
\begin{eqnarray}
\sqrt{3}RHS(\pi^{0}\Sigma^{0}\Lambda)=\frac{1}
{2\sqrt{2}}[{\cal{D}}^{(0)}
(\pi^{0};M^{2};s,d;u)+
{\cal{D}}^{(0)}(\pi^{0};M^{2};s,u;d)+
\nonumber\\
{\cal{D}}^{(0)}(\pi^{0};M^{2};u,s;d)+
{\cal{D}}^{(0)}(\pi^{0};M^{2};d,s;u)]
\rightarrow |_{ exact\quad SU(3)} \quad\sqrt{2}D .
\end{eqnarray}
Isotopic invariance 
allows to deduce the corresponding expression for
the $\pi^{-} \Sigma^{+}\bar \Lambda$ coupling:
\begin{eqnarray}
\sqrt{3}m_{\pi}^{2}\lambda_{\Lambda}\lambda_{\Sigma}
g(\pi^{-} \Sigma^{+}\bar \Lambda)
\frac{M^{2}}{M_{\Sigma}^{2}-M_{\Lambda}^{2}}
(e^{-M^{2}_{\Lambda}/M^{2}}-e^{-M_{\Sigma}^{2}/M^{2}})
[1+A_{\Sigma\Lambda}M^{2}]=
\nonumber\\
=[-m_{\pi}^{2}M^{4}E_{0}(x)[\frac{(\qu+\qd)}
{12 \pi^{2}f_{\pi}}+\frac{3f_{3\pi}}{4\sqrt{2}\pi^{2}}]+
\nonumber\\
\frac{(m_{0}^{2}+6M^{2})}{6f_{\pi}}[ (m_{u}\qs+m_{s}\qu) (\qu+\qd)]+
\nonumber\\
+\frac{m_{\pi}^{2}}{72f_{\pi}}(\qu+\qd)
\bra\frac{\alpha_{s}}{\pi}
{\cal{G}}^{2}\ket]\equiv \sqrt{2}{\cal{D}}^{(-)}(\pi^{-};M^{2};s,d;u)
\rightarrow |_{ exact\quad SU(3)} \quad\sqrt{2}D .
\label{siglampi}
\end{eqnarray}
And now we are able to derive Borel sum rules for
$K$-meson couplings to octet baryons starting from those for
$\pi\Sigma\Lambda$ and $\pi\Sigma\Sigma$ couplings
given by the Eqs.(\ref{pimisi},\ref{siglampi}).
We shall form auxiliary couplings upon using
quantities $\Lambda_{ds}, \Sigma^{0}_{ds}$ 
and $\Lambda_{us}, \Sigma^{0}_{us}$ given by the Eq.
(\ref{Lads}), 
and then return to those usual ones
performing  transformations
$d\leftrightarrow s$ and $u\leftrightarrow s$.
First we construct a formal sum rule 
for the case where $\Lambda$ is changed
to $\Lambda_{ds}$ just by using Eq.(\ref{Lads}), and we
retain for a moment only $RHS$ of the
corresponding sum rules: 
\begin{eqnarray}
RHS(\pi^{-} \Sigma^{+}\bar \Lambda_{ds})=
-\frac{\sqrt{3}}{2} RHS(\pi^{-} \Sigma^{+}{\bar \Sigma}^{0})
+\frac{1}{2}RHS(\pi^{-} \Sigma^{+}\bar \Lambda)=
\nonumber\\  
\sqrt{\frac{1}{6}}(3{\cal{F}}^{(-)}(\pi^{-};M^{2};u,d;s)+
{\cal{D}}^{(-)}(\pi^{-};M^{2};s,d;u)).
\end{eqnarray}
Performing transformation $(d\leftrightarrow s)$ we 
should change $\pi^{-}$ to $K^{-} and $ $\Sigma^{+}$ to
$-p$ to obtain:
\begin{eqnarray}
RHS((g(\pi^{-} \Sigma^{+}\bar \Lambda_{ds})_{ds})=
-RHS(g(K^{-} p \bar \Lambda )=
\nonumber\\
\sqrt{\frac{1}{6}}(3{\cal{F}}^{(-)}(K^{-};M^{2};u,s;d)+
{\cal{D}}^{(-)}(K^{-};M^{2};d,s;u))
\nonumber\\  
\rightarrow |_{ exact\quad SU(3)} 
\quad\sqrt{\frac{1}{6}}(3F+D),
\end{eqnarray}
or in full notation  
\begin{eqnarray}
m_{K}^{2}g_{K^{-} p \bar \Lambda}
\frac{\lambda_{\Lambda}\lambda_{N} M^{2}}
{(M^{2}_{\Lambda}-M^{2}_{N})}
(e^{-M^{2}_{N}/M^{2}}-e^{-M^{2}_{\Lambda}/M^{2}} )
(1+A_{\Lambda N} M^{2})
\nonumber\\  
=-\frac{1}{2\sqrt{3}}[-m_{K}^{2}M^{4}E_{0}(x)[\frac{(\qu+\qs)}
{12 \pi^{2}f_{K}}+\frac{3f_{3K}}{4\sqrt{2}\pi^{2}}]+
\nonumber\\  
\frac{(2m_{0}^{2}+3M^{2})}{3f_{K}}[ (m_{u}\qd+m_{d}\qu) (\qu+\qs)]
\nonumber\\  
+\frac{m_{K}^{2}}{72f_{K}}(\qu+\qs)
\bra\frac{\alpha_{s}}{\pi}
{\cal{G}}^{2}\ket].
\label{KpLa}
\end{eqnarray}
Interchanging $(u\leftrightarrow d)$ one 
transforms it into the sum rule for 
$g_{K^{0} n \bar \Lambda}$.

In a similar way
constructing a formal sum rule with $\Lambda_{us}$ 
we obtain:
\begin{eqnarray}
m_{K}^{2}g_{{\bar K}^{0} \Xi^{0} \bar \Lambda }
\frac{\lambda_{\Lambda}\lambda_{ \Xi} M^{2}}
{(M^{2}_{ \Xi}-M^{2}_{\Lambda})}
(e^{-M^{2}_{\Lambda}/M^{2}}-e^{-M^{2}_{ \Xi}/M^{2}} )
(1+A_{\Lambda \Xi}M^{2})=
\nonumber\\
- RHS((g(\pi^{-} \Sigma^{+}\bar \Lambda_{us})_{us})=
RHS({\bar K}^{0} \Xi^{0} \bar \Lambda )=
\nonumber\\  
\sqrt{\frac{1}{6}}(3{\cal{F}}^{(-)}(K^{0};M^{2};s,d;u)-
{\cal{D}}^{(-)}(K^{0};M^{2};u,d;s))
\nonumber\\  
\rightarrow |_{ exact\quad SU(3)} 
\quad\sqrt{\frac{1}{6}}(3F-D),
\label{KXiLa}
\end{eqnarray}
Upon interchange $(u\leftrightarrow d)$ one get the sum
rule for the coupling constant ${\bar K}^{-} \Xi^{-} \bar \Lambda$.

Analogous sum rules can be constructed for $\Sigma^{0}$
coupling with kaon. First using Eq.(\ref{Lads}) and 
Eqs.(\ref{pimisi},\ref{siglampi})
we construct RHS of the sum rule 
involving $\Sigma^{0}_{ds}$:
\begin{eqnarray}
-2 \cdot RHS(\pi^{-} \Sigma^{+}\bar \Sigma^{0}_{ds})=
RHS(\pi^{-} \Sigma^{+}\bar \Sigma^{0})+
\sqrt{3}RHS(\pi^{-} \Sigma^{+}\bar \Lambda)=
\nonumber\\
-\sqrt{2}{\cal{F}}^{(-)}(\pi^{-};M^{2};u,d;s)+
\sqrt{2}{\cal{D}}^{(-)}(\pi^{-};M^{2};s,d;u)
\end{eqnarray}
and then return to real $\Sigma^{0}$ with the 2nd 
transformation $(d\leftrightarrow s)$
changing $\Sigma^{+}$ to $-p$ and $\pi^{-} $ to $K^{-}$:
\begin{eqnarray}
2m_{K}^{2}g_{ K^{-} p \bar \Sigma^{0} }
\frac{\lambda_{\Sigma}\lambda_{ N} M^{2}}
{(M^{2}_{\Sigma}-M^{2}_{ N})}
(e^{-M^{2}_{ N}/M^{2}}-e^{-M^{2}_{\Sigma}/M^{2}} )
(1+A_{\Sigma N} M^{2})=
\nonumber\\
-2\cdot RHS((\pi^{-} \Sigma^{+}\bar \Sigma^{0}_{ds})_{ds})=
2\cdot RHS(K^{-} p \bar \Sigma^{0})=
\nonumber\\
-\sqrt{2}{\cal{F}}^{(-)}(K^{-};M^{2};u,s;d)+
\sqrt{2}{\cal{D}}^{(-)}(K^{-};M^{2};d,s;u)
\nonumber\\
\rightarrow |_{ exact\quad SU(3)} \quad-\sqrt{2}(F-D),
\label{KpSi} 
\end{eqnarray}
As the last one we construct sum rule for the
formal quantity involving $\Sigma^{0}_{us}$
to obtain finally
\begin{eqnarray}
2m_{K}^{2}g_{{\bar K}^{0} \Xi^{0} \bar \Sigma^{0} }
\frac{\lambda_{\Sigma}\lambda_{ \Xi} M^{2}}
{(M^{2}_{ \Xi}-M^{2}_{\Sigma})}
(e^{-M^{2}_{ \Sigma}/M^{2}}-e^{-M^{2}_{\Xi}/M^{2}})
(1+A_{\Sigma \Xi} M^{2})=
\nonumber\\
-\sqrt{2}{\cal{F}}^{(-)}(K^{0};M^{2};s,d;u)-
\sqrt{2}{\cal{D}}^{(-)}(K^{0};M^{2};u,d;s)
\nonumber\\
\rightarrow |_{ exact\quad SU(3)} \quad-\sqrt{2}(F+D),
\end{eqnarray}
Sum rules 
for other $g_{{\bar K} N \Sigma }$ and 
$g_{{\bar K}\Xi \Sigma}$ couplings 
are obtained with isotopic tramsformations.

\section{Summary and results}

Thus we have constructed QCD sum rules with the Lorenz structure 
$i\gamma_{5}$ for  
K meson - baryon coupling constants $g_{K N Y}$ and 
$g_{K \Xi Y}$, $Y=\Sigma,\Lambda$
starting from those for
$g_{\pi\Sigma\Sigma}$ and $g_{\pi\Sigma\Lambda}$ ones.
We have calculated 
(absolute values of) coupling constants of $K$-mesons
to octet baryons. The results are presented in the Tables 1,2.
In order to control our results we recalculate sum rules for
$\pi$ couplings to baryons obtaining values close to
those of \cite{Kim}.
As the Lorenz structure $i\gamma_{5}$ was chosen
mostly for methodical reasons 
the results do not pretend to
account for real quantities \cite{Kim}.

The sum rules confirm a known result that
unitary picture in terms of the D and F constants is not 
suitable for meson-baryon couplings due to large symmetry breaking. 
At the same time these sum rules when expressed in terms of
the generalized functions ${\cal{F}}$ and ${\cal{D}}$ 
reveal indeed a simple $SU(3)_{f}$ pattern, and this
is one of the main results we present here.
The relations obtained here
indicate in what way
one can change and use the concept of the unitary symmetry
in the framework of QCD sum rules.

\begin{theacknowledgments}
We are grateful to F.Hussain, B.L.Ioffe, G.Thompson for 
useful discussions. One of us (V.S.Z.) is grateful to
to Abdus Salam ICTP (Trieste, Italy) 
for financial support.
This work was supported in part by a Presidential grant 
N 1619.2003.2 for support of leading scientific schools.
\end{theacknowledgments}


\begin{table}
\begin{tabular}{lcccc}
\hline
\tablehead{1}{l}{b}{Coupling} &
\tablehead{1}{c}{b}{Borel Window $M^2, GeV^2$} &
\tablehead{1}{c}{b}{$g$} &
\tablehead{1}{c}{b}{$A g, GeV^{-2}$} &
\tablehead{1}{c}{b}{$A, GeV^{-2}$} \\
\hline
$\pi^{0}pp$     & 1.0-1.4 & $13.4/\sqrt{2}$ & 5.75 & 0.62  \\ \hline
$\bar K^{0}\Xi^{0}\Lambda$ 
                & 1.3-2.3 & -1.35 & 0.8 & -0.59 \\ \hline
$K^{-}p\Sigma^{0}$ & 1.1-2.1 & 1.18 & 2.53 & 2.14 \\ \hline
$\bar K^{0}\Xi^{0}\Sigma^{0}$ 
                & 1.5-2.5 & -3.09 & -0.94 & 0.30 \\ \hline
\end{tabular}
\caption{The best-fitted values of the coupling constants
$g_{KNY}$, $g_{K\Xi Y}$ and corresponding values of
$A_{NY}$, $A_{\Xi Y}$ are given together with the 
Borel windows for each sum rule, $Y=\Lambda,\Sigma$}
\label{tab:a}
\end{table}

\begin{table}
\begin{tabular}{cccc} 
\hline
\tablehead{1}{c}{b}{Coupling} & 
\tablehead{1}{c}{b}{$|g|$ \cite{Marina}} &  
\tablehead{1}{c}{b}{$g$ \cite{Choe1},\cite{Choe2}} & 
\tablehead{1}{c}{b}{$g$, this work}    \\ 
\hline
$\pi^{0}pp$   & - & - & $13.4/\sqrt{2}$(input) \\ \hline
$KN\Lambda$  & 2.37$\pm$0.09 & -3.47 & 0.77 \\ \hline
${\bar K}\Xi\Lambda$ 
                & - & -  & -1.35  \\ \hline
$KN\Sigma$  & 0.025$\pm$0.015 & 1.17 & 1.18 \\ \hline
${\bar K}\Xi\Sigma$ & - & 7.02 & -3.09 \\ \hline
\end{tabular}
\caption{The values of the coupling constants
$g_{KNY}$, $g_{K\Xi Y}$, $Y=\Lambda,\Sigma$, of this work as
well as of several recent works are given}
\end{table}

\end{document}